\documentclass[prl,twocolumn,superscriptaddress,notitlepage,nofootinbib]{revtex4-2}
\usepackage[colorlinks]{hyperref}
\hypersetup{colorlinks,linkcolor=blue,citecolor=blue,urlcolor=blue,final}
\DeclareMathAlphabet\mathbfcal{OMS}{cmsy}{b}{n}
\usepackage[utf8]{inputenc}
\usepackage[T1]{fontenc}
\usepackage[english]{babel}
\usepackage[dvipsnames]{xcolor}
\usepackage{lmodern,fancyhdr}
\usepackage{graphicx,float}
\usepackage{enumitem,amsmath,amssymb,amsthm,bm,array,mathdots,mathrsfs,dsfont,bbold}

\usepackage{lineno}

 \usepackage{url}
 \usepackage{siunitx}
 \usepackage{braket}
 \usepackage[normalem]{ulem}
\newcommand{\diff}{{\rm d}}

\newcommand{\toramp}{\varphi}
\newcommand{\polangle}{\theta}
\newcommand{\myeqref}{Eq.~\eqref}

\newcommand{\commentOut}[1]{}
\usepackage{scalefnt}   

\newcommand{\affil}{Department of Physics, Humboldt-Universität zu Berlin, 10099 Berlin, Germany}
\newcommand{\affilvienna}{Vienna Center for Quantum Science and Technology, TU Wien-Atominstitut, Stadionallee 2, 1020 Vienna, Austria}
\newcommand{\affilpresentaddressmichael}{Present address: University of Vienna, Faculty of Physics, Vienna Center for Quantum Science and Technology, Boltzmanngasse 5, 1090 Vienna, Austria}
\newcommand{\affilpresenaddresssebastian}{Present address: Max-Planck-Institut f{\"u}r Quantenoptik, Hans-Kopfermann-Straße 1, 85748 Garching, Germany}

\begin{document}
\scalefont{1.05}
\title{Feedback–cooling the fundamental torsional mechanical mode of a tapered optical fiber to 30 mK}

\author{Felix Tebbenjohanns}
\affiliation{\affil}
\author{Jihao Jia}
\affiliation{\affil}
\author{Michael Antesberger}
\affiliation{\affilvienna}
\affiliation{\affilpresentaddressmichael}
\author{Adarsh Shankar Prasad}
\affiliation{\affilvienna}
\author{Sebastian Pucher}
\affiliation{\affil}
\affiliation{\affilpresenaddresssebastian}
\author{Arno Rauschenbeutel}
\affiliation{\affil}
\affiliation{\affilvienna}
\author{J{\"u}rgen Volz}
\affiliation{\affil}
\affiliation{\affilvienna}
\author{Philipp Schneeweiss}
\email{philipp.schneeweiss@hu-berlin.de}
\affiliation{\affil}
\affiliation{\affilvienna}

\begin{abstract}
We experimentally 
study the fundamental torsional mechanical mode of the nanofiber waist of a tapered optical fiber. 
We find that this oscillator with a resonance frequency of $f=161.7$~kHz features a quality factor of up to $Q=10^7$ and a $Q$-frequency product of $Qf=1$~THz. 
The polarization fluctuations of a transmitted laser field serve as a probe for the torsional motion. 
We damp the oscillator's thermal motion from room temperature to 28(7)~mK by means of active feedback.
Our results might enable new types of fiber--based sensors and lay the foundation for a novel hybrid quantum optomechanical platform.
\end{abstract}

\date\today

\maketitle

Tapered optical fibers (TOFs) are versatile tools~\cite{Tong2012} with applications ranging from supercontinuum generation~\cite{Birks2000} to optical sensors~\cite{Warken2007,Brambilla2010} and high-power fiber amplifiers~\cite{Filippov2008} to optical interfaces for atoms~\cite{Nayak2007,Sague2007, Morrissey2013}. 
The mechanical excitations of TOFs have been studied in the context of Brillouin scattering~\cite{Beugnot2014} and for the development of acousto-optical devices~\cite{Birks1996}.
In the case of TOFs with a sub-wavelength-diameter ``nanofiber'' waist, however, mechanical motion is typically considered a nuisance as it leads, e.g., to heating in nanofiber--based cold--atom traps~\cite{Vetsch2010,Huemmer19}.
Such nanofibers feature a particular mode of torsional motion, which is well confined to the waist, forming a mechanical resonator with reported quality factors just below $Q=\SI{e5}{}$~\cite{Wuttke13Optically, Fenton18, Su22}.
At the same time, this torsional motion couples to the polarization of the transmitted light field, which enabled, e.g., active cooling of the oscillator's thermal motion by a factor of five~\cite{Su22}.
In addition, cooling of flexural modes of a TOF has been demonstrated~\cite{Pennetta2020}.
These advances lead to the question of whether the mechanical motion of nanofibers could be used in the field of quantum optomechanics, where one requires an extraordinary mechanical decoupling from the environment. 
In particular, it is desirable that the product $Qf=\Omega^2/(2\pi\gamma_0)$ exceeds $k_B T_0/h=\SI{6}{THz}$ (with the angular resonance frequency $\Omega$, the intrinsic damping rate $\gamma_0$, room temperature $T_0$, Boltzmann's constant $k_B$, and Planck's constant $h$)~\cite{Aspelmeyer2014}.
In various optomechanical platforms, this requirement has recently led to impressive progress~\cite{Tsaturyan2017,Bereyhi2022Perimeter} with reported  $Qf$ products reaching \SI{e4}{THz}~\cite{Beccari2022}.
However, most clamped optomechanical systems rely on a high-order mechanical mode of the structure with the resonance inside a narrow phononic bandgap. Since this can be a drawback for quantum optomechanical experiments, sophisticated structures have been fabricated, which achieve ultra-low dissipation of the fundamental mechanical mode with $Qf$ exceeding \SI{10}{THz}~\cite{Hoj2021,Bereyhi2022Structural,Pratt2023}. 
For fundamental modes of motion, these values are only surpassed by optically levitated silica particles, whose damping is given by the vacuum pressure~\cite{Gieseler2012,Magrini2021}.
A prerequisite for any quantum protocol with mechanical oscillators is their preparation close to the motional ground state, 
which has been achieved with a variety of mechanical oscillators using different techniques~\cite{OConnell2010, Chan2011, Teufel2011}, including measurement-based feedback cooling~\cite{Rossi2018}.

Here, we experimentally investigate the fundamental mode of torsional motion of a nanofiber in vacuum and achieve a quality factor up to $Q=\SI{e7}{}$ and a $Qf$ product exceeding \SI{1}{THz}. This represents a hundred-fold improvement of the reported mechanical quality factor of a TOF.
We optically measure the torsional displacement of the nanofiber and show that it equilibrates at room temperature.
Finally, we use measurement-based feedback to cool the motion by four orders of magnitude to a temperature of \SI{28(7)}{mK}, thereby demonstrating that such an optical nanofiber represents a competitive optomechanical platform.

\begin{figure}[!t]
\centering
{\includegraphics[width=1.0\linewidth]{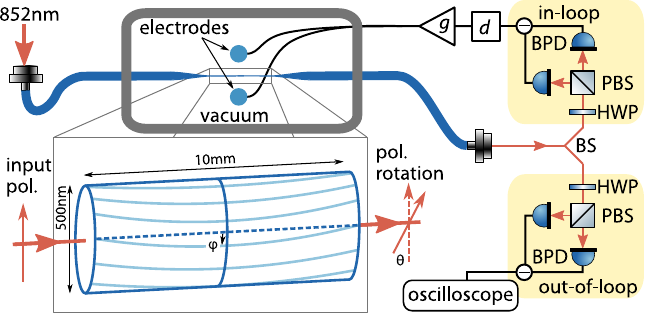}}
\caption{Schematic of the experimental setup.  We launch a linearly polarized probe light field through an optical nanofiber that is placed inside a vacuum chamber. 
The mechanical torsional mode is illustrated in the bottom left. 
We split the transmitted probe light and perform an out-of-loop analysis of the polarization rotation, $\polangle$, induced by the angular displacement of the nanofiber, $\toramp$. We use a second, in-loop analysis setup and feedback to two electrodes below and above the nanofiber to cool its torsional motion. Balanced photodiode (BPD), half-wave plate (HWP), polarizing beamsplitter (PBS), 50:50 beamsplitter (BS), variable delay $d$, and feedback gain $g$.
}
\label{fig:setup}
\end{figure}

\begin{figure}[!t]
\centering
{\includegraphics[width=\linewidth]{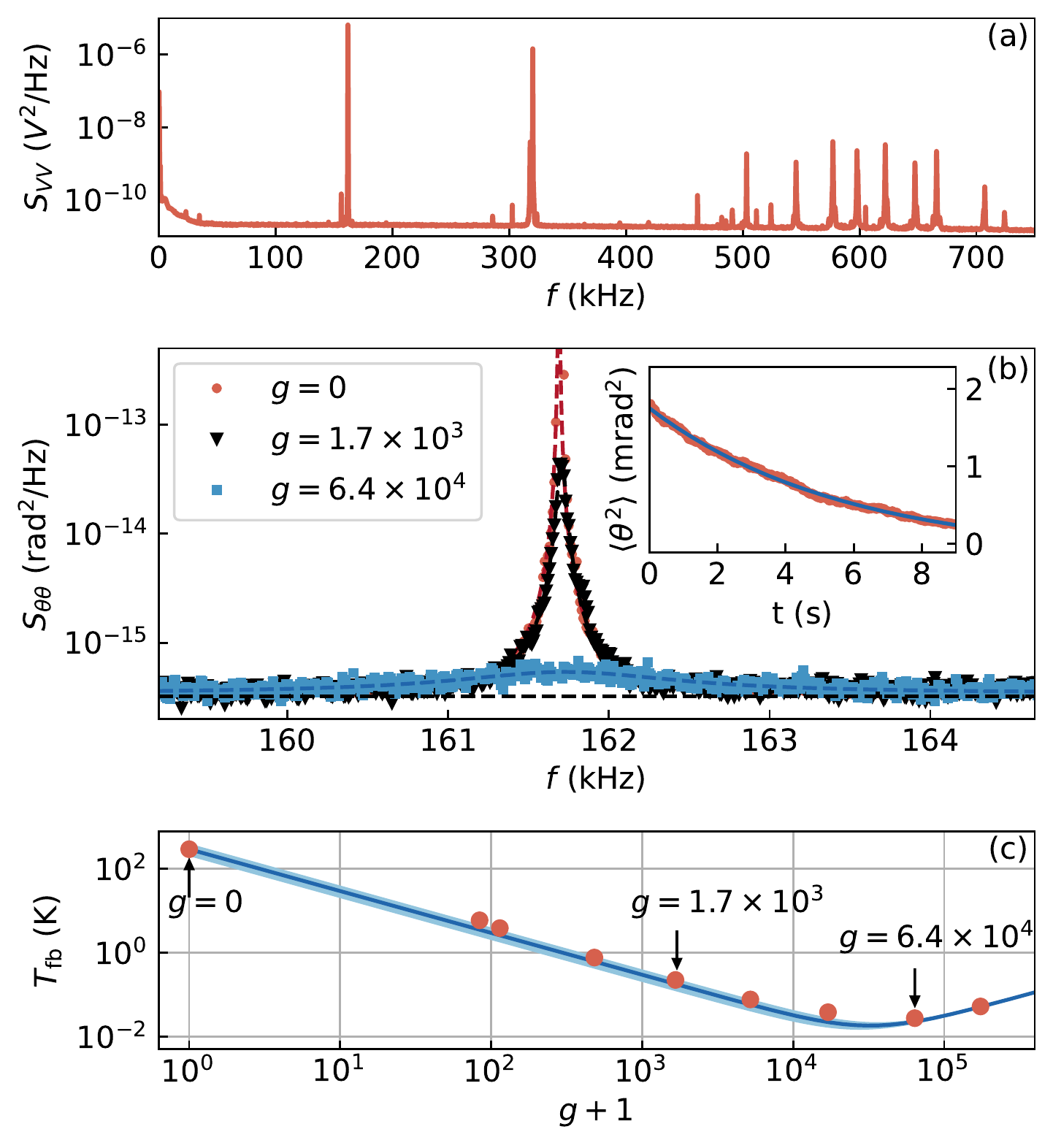}}
\caption{Characterization and cooling of the thermally excited torsional motion. 
(a) Power spectral density (PSD), $S_{VV}(f)$, of the out-of-loop analyzer voltage with \SI{100}{Hz} resolution.
The prominent peaks at $f=\SI{161.7}{kHz}$ and $f=\SI{319.8}{kHz}$ correspond to the fundamental and second harmonic mode of the torsional motion, respectively.
(b) PSD of the polarization rotation, $S_{\polangle\polangle}(f)$, with \SI{10}{Hz} resolution, zoomed in around the resonance frequency of the fundamental torsional mode in equilibrium with room temperature (red circles), and cooled with a moderate  and strong feedback gain, $g$, (black triangles and dark blue squares, respectively).
The dashed lines are Lorentzian fits.
The detection noise floor (black dashed) is estimated from the measured noise around the Lorentzian peaks.
Inset: ring-down measurement of the fundamental mode at a vacuum pressure of \SI{2e-7}{mbar}, revealing a damping rate of  $\gamma/(2\pi)=\SI{28(1)}{mHz}$ and a quality factor of $Q= \SI{5.8(1)e6}{}$.
(c) Effective temperature, $T_\text{fb}$, under feedback as a function of the feedback gain $g$. By increasing the feedback gain from $g=0$ to $g=\SI{6.4e4}{}$, 
we cool the motion by four orders of magnitude from room temperature to \SI{28(7)}{mK}.
The blue line is a model according to \myeqref{eq:feedback_energy}.
Its thickness represents the statistical error of 26\% of the motional energy measured at $g= 0$. This error is because of a finite measurement time of $\mathcal{T} = \SI{160}{s}$, as we detail in the main text.
}
\label{fig:psds}
\end{figure}

Our experimental setup is sketched in Fig.~\ref{fig:setup}. 
We mount a TOF with a nanofiber waist in a vacuum chamber with a base pressure of \SI{2e-7}{mbar}.  
The TOF is manufactured from a commercial single-mode fiber (Thorlabs SM800-5.6-125) in a heat-and-pull process and features a homogeneous waist diameter of $2R=\SI{500}{nm}$ over a length of $L=\SI{10}{mm}$~\cite{Birks1992}. In the process fixing the TOF on the fiber holder, it is slightly stretched, however, by much less that its rupture strain~\cite{Holleis2014}.
We transmit linearly polarized laser light with a wavelength of \SI{852}{nm} through the nanofiber. After mounting it in the vacuum chamber and pumping down, we increase the optical power of this light to about \SI{20}{mW} before lowering it to \SI{1}{mW}. This procedure considerably decreases the mechanical damping rate of the torsional fiber motion.
Two identical setups for polarization analysis, which are based on a half-wave plate, a polarizing beamsplitter, and a balanced photodiode, measure the polarization fluctuations of the transmitted light.
We verified that the noise floor of both setups is dominated by photon shot noise. 
The first ``out-of-loop'' analysis setup is used to characterize the motion, while feedback from the second ``in-loop'' analysis setup is later used to cool the nanofiber motion.
In Fig.~\ref{fig:psds}(a), we present the power spectral density (PSD), $S_{VV}(f)$, of the out-of-loop analyzer signal recorded with an oscilloscope over a frequency range of \SI{750}{kHz}, while the system is in equilibrium with its environment.
The various peaks in the PSD have been identified as different torsional modes of the nanofiber and TOF motion in Ref.~\cite{Wuttke13Optically}. In particular, the two dominant peaks at frequencies $f=\SI{161.7}{kHz}$ and $f=\SI{319.8}{kHz}$ correspond to the fundamental and second harmonic torsional mode of the nanofiber.
For these modes, the tapers act as mirrors such that they form standing waves of the nanofiber's angular displacement, as depicted in the inset of Fig.~\ref{fig:setup}, and constitute mechanical harmonic oscillators. 
While in this work we focus on the  fundamental mode, we note that the peaks above $f=\SI{500}{kHz}$ correspond to modes with resonances above the cut-off frequency of the taper, thus involving the entire TOF structure.
Due to birefringence of the nanofiber, possibly originating from an elliptical cross-section after fabrication, the fiber's angular displacement, $\toramp$, leads to a rotation of the polarization of the transmitted light, $\polangle$, which we deduce from the analyzer signal using Malus' law.
In Fig.~\ref{fig:psds}(b), we show as red datapoints a zoom into the PSD, $S_{\polangle\polangle}(f)$, around the resonance frequency of the fundamental torsional mode with a resolution of \SI{10}{Hz}, and a Lorentzian fit (red dashed line). 
The dynamics of $\toramp$ follows the equation of motion of a harmonic oscillator (with $|\toramp|\ll1$)
\begin{equation}
    \ddot{\toramp} + \gamma\dot \toramp + \Omega^2\toramp = [M_\text{th}(t)+M_\text{ext}(t)]/I_\text{eff}\,,
\end{equation}
where $\gamma$ is the energy damping rate, $\Omega/(2\pi) = \SI{161.7}{kHz}$ is the eigenfrequency, and $I_\text{eff}$ is the effective moment of inertia of the fundamental torsional mode. 
For the fundamental torsional mode, $I_\text{eff}=(\pi/4)\rho LR^4=\SI{6.7e-26}{kgm^2}$ is given by half the value of a rigid cylinder of the same length and radius rotating about its symmetry axis.
The torque acting on the fiber has a white--noise contribution $M_\text{th}(t)$ with $\braket{M_\text{th}(t)}=0$ and $\braket{M_\text{th}(t)M_\text{th}(t')}=2I_\text{eff}\gamma k_B T_0\delta(t-t')$, such that the motional energy equilibrates with its bath at room temperature $T_0$. 
In addition, we exert an external torque, $M_\text{ext}(t)$, by applying a sinusoidal voltage, $V(t)$, with moderate amplitude of up to \SI{20}{V} across two electrodes, which sandwich the nanofiber and which are separated by about \SI{5}{mm}, see Fig.~\ref{fig:setup}~\cite{Wuttke13Optically}.
By measuring the amplitude and phase response of the overall system, we can confirm that it can be well described by a driven harmonic oscillator, see our supplemental material for a Bode plot~\cite{SI}.
This cross check proves the linearity of both our actuator [$M_\text{ext}(t)\propto V(t)$] and our analyzer signal [$\polangle(t)\propto \toramp(t)$].
The exact mechanism of how the applied electric field exerts a torque on the nanofiber is not yet fully understood.
We hypothesize that the interaction is due to a Coulomb force acting on stray charges on the nanofiber's surface. From experiments with optically levitated particles, which employ such a Coulomb interaction, it is known that charges on nanoscale silica objects can stay for many days~\cite{Frimmer2017}. Indeed, we did not observe any change in the interaction strength during the measurements presented in this work.

\begin{figure}[!t]
\centering
{\includegraphics[width=\linewidth]{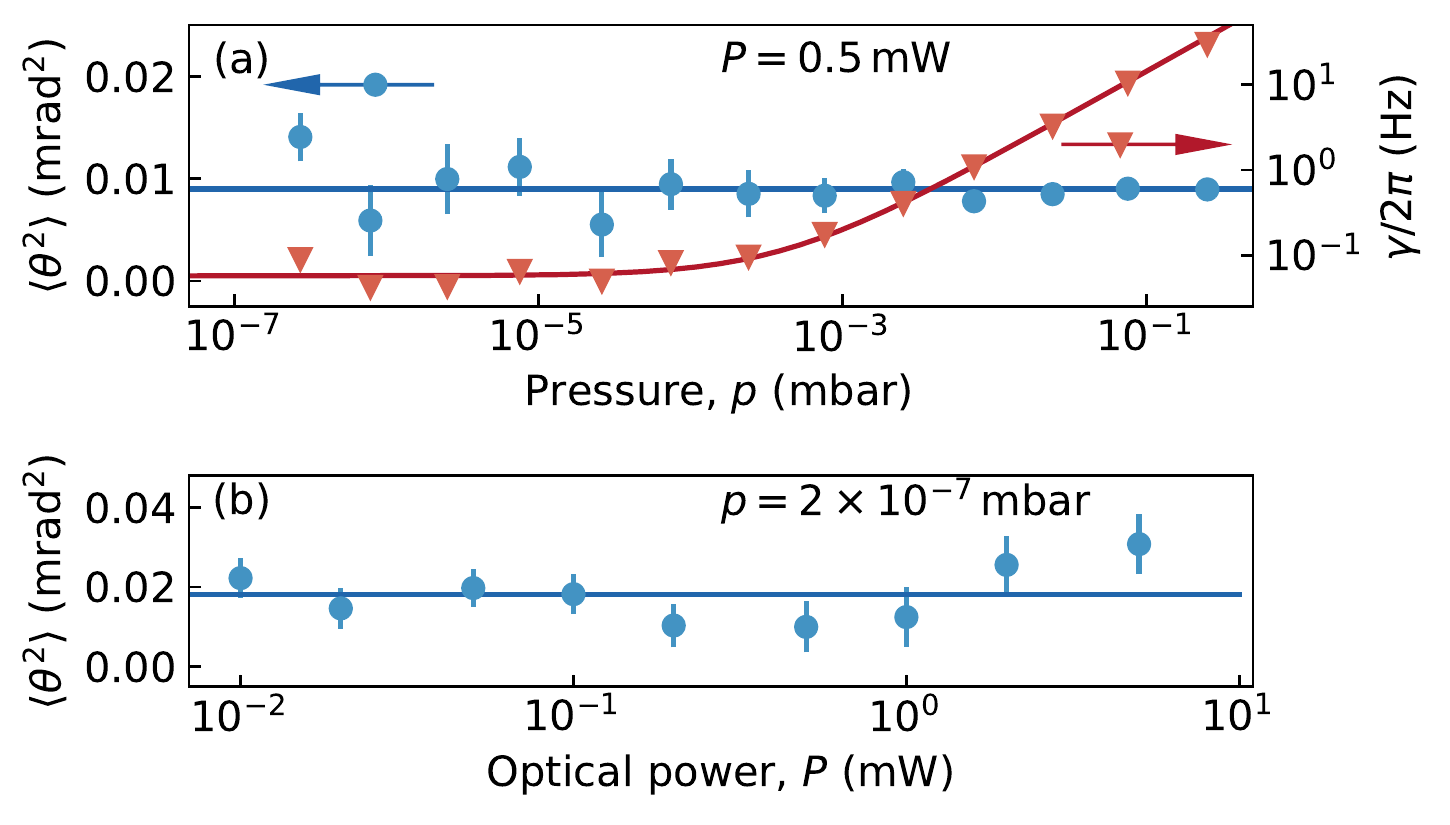}}
\caption{Pressure and power dependency of the motional energy. (a) Detected signal power $\braket{\polangle^2}$ (left axis, blue dots) and damping rate $\gamma$ (right axis, red triangles) as a function of gas pressure $p$ in the vacuum chamber. The optical power in the fiber is $P=\SI{0.5}{mW}$. Due to a finite measurement time of $\mathcal T=\SI{50}{s}$, the observed energy fluctuates, as indicated by error bars and discussed in the main text.
The blue solid line indicates the average value of the data points, revealing that the motional energy is independent of the vacuum pressure. 
The red solid line is a fit of the form $\gamma = \gamma_0 + ap$. The extracted intrinsic damping rate of the torsional motion is $\gamma_0/(2\pi)=\SI{57(5)}{mHz}$.
(b) Signal power as a function of the optical power $P$ in the nanofiber at a fixed vacuum pressure of \SI{2e-7}{mbar}. 
The solid line indicates the average signal power of the data points, revealing that the motional energy is virtually independent of the optical power in the fiber.
}
\label{fig:dependence}
\end{figure}

Let us now characterize the damping rate $\gamma$. 
The fundamental peak at  $\Omega/(2\pi)= \SI{161.7}{kHz}$ has a full-width at half maximum of about \SI{1}{Hz}, which we extract from PSD data with a Fourier-limited frequency resolution of \SI{6.25}{mHz} (not shown).
It turns out that this peak is broadened by drifts of the resonance frequency at the \SI{1}{Hz}-level, such that we cannot accurately infer $\gamma$ from a Lorentzian fit. Instead,
we perform a ring-down measurement by resonantly exciting the torsional motion via a voltage applied to the electrodes. After switching off the drive, we record the polarization fluctuations $\polangle(t)$ with the oscilloscope.
We numerically perform frequency demodulation of the recorded signal at the eigenfrequency $\Omega$ in post processing using a $2^{\rm nd}$-order Butterworth filter with a bandwidth of \SI{1}{kHz}.
The squared modulus of the resulting complex--valued oscillation amplitude represents an energy ring-down measurement and is shown in the inset of Fig.~\ref{fig:psds}(b).
By fitting an exponential decay, we find that our mechanical oscillator has a damping rate of $\gamma/(2\pi)= \SI{28(1)}{mHz}$ and thus a quality factor of $Q=\Omega/\gamma = \SI{5.8(1)e6}{}$.
For five different nanofibers with the same nominal diameter and length, we repeatedly found large $Q$ factors exceeding $Q=\SI{5e6}{}$ and up to $Q=\SI{e7}{}$. This is more than two orders of magnitude larger than the highest value that has been reported so far~\cite{Su22}.
In fact, after mounting our nanofiber inside the vacuum chamber and pumping down, we initially measure a moderate $Q$ factor of around \SI{e4}{}, in agreement with Refs.~\cite{Wuttke13Optically, Fenton18, Su22}. The $Q$ factor increases to the reported value only after increasing the TOF-guided optical power to about \SI{20}{mW} for a few seconds and then reducing it again.
The $Q$ factor maintains its high value after this preparation procedure and then drops slightly over the course of a day.
We observe that also the $Q$ factor of the second harmonic increases, to about $10^6$. The dependence of the mechanical Q factor on strain and a possible further increase of Q due to dissipation dilution~\cite{Pratt2023} is a matter of further research.
In Fig.~\ref{fig:dependence}(a), we measure $\gamma$ as a function of the vacuum pressure, $p$, (red triangles). 
For pressures above \SI{e-4}{mbar}, $\gamma$ increases linearly with the pressure, which means that the motion is damped by the residual gas~\cite{Wuttke13Optically}. 
By fitting a function of the form $\gamma = \gamma_0+a p$ to the data we find the intrinsic damping rate of $\gamma_0/(2\pi) = \SI{57(5)}{mHz}$.
This damping rate is larger than the one reported above due to the fact that we performed this measurement some days after the preparation procedure so that the quality factor had decreased.

Having established both the measurement and the actuation of the torsional motion, we now close the loop using a feedback filter to cool the motion, i.e., we perform ``cold damping''~\cite{Mancini1998, Cohadon1999, Bushev2006, Kleckner2006, Rossi2018}.
For this, we delay the signal $\polangle(t)\propto\toramp(t)$ from our in-loop analyzer by a variable delay $d$, amplify it with a variable gain $g$, and apply feedback in the form of a torque $M_\text{ext}(t) = g \gamma_0 \Omega I_\text{eff} \toramp(t-d)$, see Fig.~\ref{fig:setup}.
Using digital electronics, we vary $d$ in steps of \SI{32}{ns} until the feedback dampens the motion. 
At this point, the feedback torque at resonance is \SI{90}{\degree} out of phase with the angular displacement $\toramp(t)$, and therefore $M_\text{ext}(t)$ is proportional to the angular velocity $\dot\toramp(t)$. In Fig.~\ref{fig:psds}(b), we show the PSD observed with the out-of-loop polarization analyzer when $g$ is set to maximal cooling performance (dark blue squares). We fit the feedback--cooled signal with a Lorentzian and
use its width, $\gamma_\text{fb}$, to infer the feedback gain $g=\SI{6.4(4)e4}{}$ via $\gamma_\text{fb}=(g+1)\gamma_0$. For comparison, we also show an intermediate value of $g=\SI{1.7(1)e3}{}$ (black triangles).
In Fig.~\ref{fig:psds}(c), we plot the effective temperature of the motion as a function of $g$.
For this, we integrate the out-of-loop PSDs in a frequency band of \SI{5}{kHz} around the resonance after subtracting the noise floor. 
We calibrate this integrated area to an effective temperature, $T_\text{fb}$, by assuming that in the absence of feedback ($g=0$), the motion equilibrates at room temperature with $T_0=\SI{295}{K}$.
We justify this assumption further below. 
At $g=\SI{6.4(4)e4}{}$ we find an effective temperature of \SI{28(7)}{mK}.
The \SI{26}{\%} relative statistical error is due to the finite measurement time of $\mathcal{T}=\SI{160}{s}$ during the calibration with $g=0$ (see below). 
At even larger feedback gains, the detection noise is fed back and heats up the motion. 
In that regime, we observe so-called noise squashing in the in-loop signal, because of the correlations between the torsional motion and the measurement noise~\cite{Poggio2007, Wilson2015}.
The blue line is a model of the theoretically expected temperature dependence on $g$ according to~\cite{Poggio2007, Tebbenjohanns2019}
\begin{equation}
\label{eq:feedback_energy}
    T_\text{fb} =  T_0\left(\frac{1}{1+g} + \frac{g}{\rm SNR}\right).
\end{equation}
Here, we assume that the system operates in the underdamped regime even with feedback, i.e., $\gamma_\text{fb}\ll\Omega$.
We stress that we do not find the parameter signal-to-noise ratio (SNR) by fitting \myeqref{eq:feedback_energy} to our data in Fig.~\ref{fig:psds}(c). Instead, we independently estimate the SNR of the in-loop detector in the absence of feedback (at $g=0$), and find ${\rm SNR}=4\braket{\polangle^2}/(\gamma S_{\polangle\polangle}^{\rm noise})=\SI{1.0e9}{}$ 
(with the measured signal power $\braket{\polangle^2}$, the damping rate $\gamma$, and the measurement noise floor $S_{\polangle\polangle}^{\rm noise}$~\footnote{We define the single-sided PSD $S_{xx}(f)$ of a signal $x$ such that $\int_0^\infty \diff f~S_{xx}(f) = \braket{x^2}$.}).
The agreement of model and data indicates that our feedback is close to ideal given the measurement noise.

Finally, let us estimate the effective temperature of the mechanical oscillator in the absence of feedback, which possibly differs from our laboratory's room temperature for at least two reasons.
Firstly, there might be a source of (mechanical) noise that resonantly increases the energy of the mode, and secondly, the TOF-guided optical power might increase the temperature of the nanofiber~\cite{Wuttke13Thermalization}.
In order to exclude these effects, we study the signal power $\braket{\polangle^2}$ as a function of the vacuum pressure $p$ while $P=\SI{0.5}{mW}$ [in Fig.~\ref{fig:dependence}(a)], and as a function of the optical power $P$ while $p=\SI{2e-7}{mbar}$ [in Fig.~\ref{fig:dependence}(b)].
To estimate $\braket{\polangle^2}$, we integrate the PSD in a bandwidth of \SI{5}{kHz} around the resonance frequency.
The error bars represent the expected energy fluctuations due to the finite measurement time of $\mathcal{T}=\SI{50}{s}$ and are given by $\sigma = \sqrt{2/(\gamma \mathcal T)} \braket{\polangle^2}$~\cite{FrenkelDaan2001, Hebestreit2018}.
Within this error, the observed signal power is largely independent of both $p$ and $P$. 
The effective temperature of the motion in the absence of feedback is thus about constant throughout the parameter space studied here.
Moreover, for pressures above \SI{e-3}{mbar}, $\gamma$ is proportional to $p$ while $\braket{\polangle^2}$ is a constant, which indicates that the predominant damping and heating mechanisms originate from the background gas in that pressure range. 
Therefore, we can conclude that, independently of the pressure, our oscillator is indeed equilibrated at the gas temperature of \SI{295}{K}.
For pressures below about \SI{e-4}{mbar}, the damping rate becomes independent of the gas pressure.
There, the residual damping might originate from clamping losses in the tapers, from surface loss, or from intrinsic volume losses~\cite{Villanueva2014}.

In summary, we have shown that the fundamental torsional motion of the nanofiber waist of a standard tapered optical fiber can exhibit an unexpectedly high quality factor of up to \SI{e7}{} and a $Qf$ product of \SI{1}{THz} without much experimental overhead.
In particular, the $Qf$ product is promisingly close to the required value of $k_BT_0/h=\SI{6}{THz}$ for ground--state cooling starting from room temperature $T_0$~\cite{Aspelmeyer2014, Tsaturyan2017, Hoj2021} and close to the largest value reported for fundamental mechanical modes.
We have shown that the system equilibrates at room temperature even at low vacuum pressures and when probed with an optical power as high as \SI{1}{mW}.
In addition, we used feedback cooling to reduce the thermal motion from room temperature by a factor of about $10^4$ to \SI{28(7)}{mK}.
This corresponds to a phonon occupation of $\bar n = k_B T/(h f) = 3.6(9)\times10^3$.
In an independent study that has been performed in parallel to ours, similar results have been obtained with all-optical feedback~\cite{Su2023}. There, the torsional mode's Q factor is about $10^5$ and a cooling by a factor of $600$ is reported. 

In the future, cooling to lower temperatures could be achieved by placing the TOF in a colder environment, e.g., in a commonly available 4K cryostat~\cite{Huetner2020}, where $k_B T_0/h$ is only \SI{83}{GHz}. In this case, our $Qf$ product would be sufficiently large for ground--state cooling,
provided that the mechanical properties are not altered and that the TOF thermalizes at this temperature.
Another important aspect is a high enough optomechanical interaction strength, which can be significantly enhanced by placing the nanofiber inside an optical cavity. 
Optical resonators that contain a nanofiber section have already been demonstrated~\cite{Wuttke13Optically, Ruddell2017, Schneeweiss2017}, including with a finesse of more than 1000~\cite{Ruddell2020}.
Finally, given the fact that thousands of laser-cooled atoms can be trapped in the evanescent field of such a nanofiber and can be controlled exceptionally well, both in their motional~\cite{Meng2018} and internal degrees of freedom~\cite{Mitsch2014}, we envision a new hybrid quantum system~\cite{Treutlein2014} where the mechanical motion of a nanofiber is coupled to the external and internal states of nanofiber--trapped atoms. 
A coherent coupling between the azimuthal motion of such nanofiber--trapped atoms with their spin degree of freedom has already been demonstrated~\cite{Dareau2018}.

\begin{acknowledgments}
We thank Thomas Hoinkes for fabricating our nanofibers, Fabrice von Chamier-Gliszczynski for the construction of the fiber mount, and Constanze Bach, Martin Frimmer, and Riccardo Pennetta for insightful discussions.
We acknowledge financial support by the Alexander von Humboldt Foundation in the framework of an Alexander von Humboldt Professorship endowed by the Federal Ministry of Education and Research, as well as funding by the Austrian Science Fund (DK CoQuS Project No.~W 1210-N16).
\end{acknowledgments}

\bibliography{main}

\end{document}


\scalefont{1.05}
\title{Supplemental Material\\Feedback–cooling the fundamental torsional mechanical mode of a tapered optical fiber to 30 mK}

\author{Felix Tebbenjohanns}
\affiliation{\affil}
\author{Jihao Jia}
\affiliation{\affil}
\author{Michael Antesberger}
\affiliation{\affilvienna}
\affiliation{\affilpresentaddressmichael}
\author{Adarsh Shankar Prasad}
\affiliation{\affilvienna}
\author{Sebastian Pucher}
\affiliation{\affil}
\affiliation{\affilpresenaddresssebastian}
\author{Arno Rauschenbeutel}
\affiliation{\affil}
\affiliation{\affilvienna}
\author{J{\"u}rgen Volz}
\affiliation{\affil}
\affiliation{\affilvienna}
\author{Philipp Schneeweiss}
\email{philipp.schneeweiss@hu-berlin.de}
\affiliation{\affil}
\affiliation{\affilvienna}

\date\today

\maketitle

\section{Amplitude and phase response}

\begin{figure}
\centering
{\includegraphics[width=0.7\linewidth]{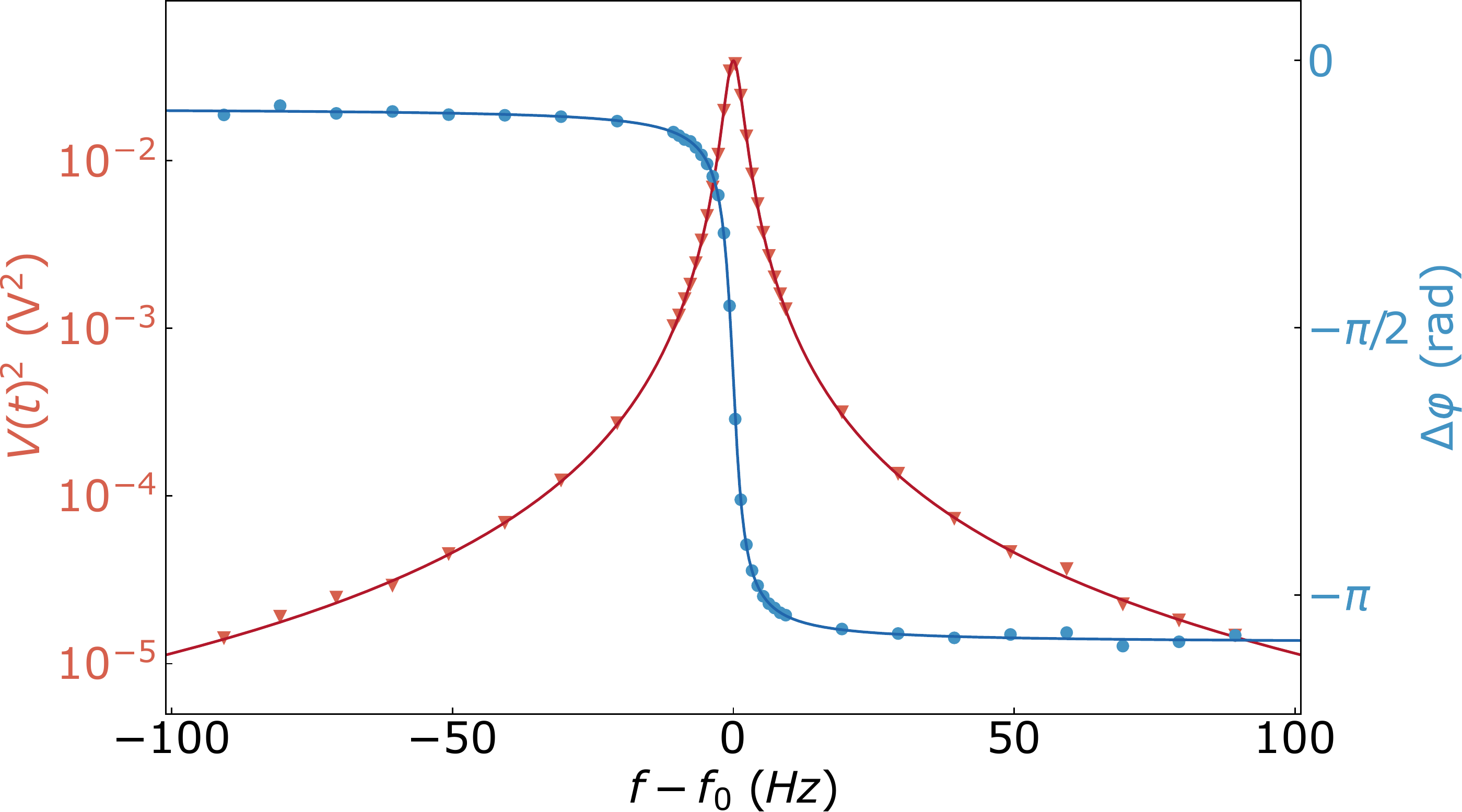}}
\caption{Bode plot of the overall system. We show the phase and the square of the amplitude of the voltage on the balanced photodetector as a function of the frequency at which the fundamental torsional mode of our tapered optical fiber is driven.
}
\label{fig:bode_plot}
\end{figure}

In the main text, we explain how we drive the torsional mechanical mode using an electric field applied across the fiber and how we measure the angular displacement of the torsional mode by analyzing the polarization of a transmitted light field. 
We apply a periodically varying voltage described by a sinusoidal function of the type $U(t)=U_0 \cos(2\pi f t)$ with $U_0 = \SI{20}{V}$ to the two electrodes below and above the nanofiber. We then measure the amplitude and phase response of the voltage detected on a balanced photodiode and described by $V(t)=V(f)\cos(2\pi f t  + \Delta \varphi(f))$ as a function of the drive frequency $f$.
The result is shown in Fig.~\ref{fig:bode_plot}. 
We note that this data has been taken with a different nanofiber than the data shown in the main manuscript. However, the shape (radius, length, taper profile) of the tapered fibers is nominally identical.
The amplitude response, $V(f)^2$, is shown as red triangles, while the phase response, $\Delta \varphi(f)$ is shown as blue dots. The solid lines are a fit of a Lorentzian function with an additional constant phase, $\varphi_0$, i.e.
\begin{subequations}
\begin{align}
V(f)^2 &= V_0^2\frac{\gamma^2/4}{(2\pi)^2(f-f_0)^2 + \gamma^2/4}, \\
\Delta \varphi(f) &= \varphi_0 - \frac{\pi}{2} - \arctan\left(\frac{4\pi(f-f_0)}{\gamma}\right).
\end{align}
\end{subequations}
The agreement between a Lorentzian model and our data shows that our system is linear both in the actuation (the applied torque) and in the detection (the polarization detection). 
From the data, we can extract the resonance frequency $f_0=\SI{162.9}{kHz}$, the damping rate $\gamma = 2\pi\times \SI{3}{Hz}$. We performed this measurement when the torsional mode was in the ``low-Q'' regime, i.e. $Q=\SI{5.4e4}{}$.
We attribute the constant phase of $\varphi_0 = \SI{16}{\degree}$ to the overall delay and phase response of the electronic components such as the photodetector in the loop.

